# HYBRID RECOMMENDER SYSTEM BASED ON PERSONAL BEHAVIOR MINING

Zhiyuan Fang, Lingqi Zhang, Kun Chen
Department of Electronic and Electrical Engineering
Department of Computer Science and Engineering
South University of Science and Technology of China
Shenzhen, China
e-mail: fangzy@mail.sustc.edu.cn
zhanglq@mail.sustc.edu.cn

## Abstract

Recommender systems are mostly well known for their applications in e-commerce sites and are mostly static models. Classical personalized recommender algorithm includes item-based collaborative filtering method applied in Amazon, matrix factorization based collaborative filtering algorithm from Netflix, etc. In this article, we hope to combine traditional model with behaviour pattern extraction method. We use desensitized mobile transaction record provided by T-mall, Alibaba to build a hybrid dynamic recommender system. The sequential pattern mining aims to find frequent sequential pattern in sequence database and is applied in this hybrid model to predict customers' payment behaviour thus contributing to the accuracy of the model.

## INTRODUCTION

Recommender systems are ubiquitous.   There perform quite well in recommending products, movies, music, etc. Traditional methods of recommender systems include content based recommender system which make recommendations that is similar to what the user has purchased in the past. While the key part of the content-based recommender system is that this model requires a feature extraction process. Take NewsWeeder [19] an example, this system in 1995 solve a significant problem in information filtering systems that is the creation of a user profile that describes the user's interests. The NewsWeeder's way of creating such profiles have been applied in content-based recommender system, which recommend a product to a customer based on a description of the item and a profile of the user's interests. But the shortcoming of this system is that in most cases only a very shallow analysis of certain kinds of content can be supplied [18], especially in some e-commerce sites because of the privacy-preserving policy.

As for CF (collaborative filtering) recommendation, the system will not recommend items that is similar to what this user has purchased, but the items what other users have liked. Collaborative filtering requires no profile or content of both users or items, recommendations are make solely based on similarity matching to other users (the nearest neighbours' searching). Traditional CF method include memory based and model based Collaborative Filtering algorithm. Matrix factorization based Collaborative Filtering algorithm are proposed by Yehuda Koren [8] in 2009, allowing the incorporation of additional information such as implicit feedback, temporal effect and etc.

Most of those recommendation models are static, in the sense they use static features to describe products and users. But nowadays people are using mobile devices to browse products everywhere, generating a lot more behavioural data. More importantly, as the density of behavioural data growth, there may be a potential chance to make use of the behavioural pattern of users. We proposed a hybrid recommender system that combine the prefix span algorithm with traditional matrix factorization.

**Corresponding author:**
Zhiyuan Fang, Department of Computer Science, South University of Science and Technology, Shenzhen, 518055, China
Email: fangzy@sustc.edu.cn

In the literature review, we will briefly overview these two algorithms. Then we will give a detailed explanation of how our system works, followed by experimenting results. Recommender system will evolve as user's behavioural data with explicit sequential pattern become abundant, and we think hybrid model is a great approach.

## LITERATURE REVIEW AND METHODOLOGY

### *1.1. Recommender System*

Traditional recommender systems like the content based recommender systems are originally derived from text documents recommendations where information retrieval technique plays an important role in extracting the features of the documents [18]. Such system tries to recommend items similar to what a customer has purchased in the past. Besides that, content based recommender system requires a comparison of users' profiles with descriptions of items, on which the recommendation will be based.
Considering the hybrid systems exist both in industry and academia which combine content based systems and collaborative filtering systems which will be discussed in the later section, we first give a definition that a pure content based system is the one that will make the recommendations based solely on the content of items rated by users. Content based systems analyse the content or descriptions of items that are rated by the user, the details of such system differs because of the data representing [19], so we firstly give a generalized input data matrix in Table 1 and Table 2.
In Table 1, the 'Song-Users matrix' include four songs ,4 users and their rated information. Number 0, 1, 2, 3 in the matrix represent the user's feedback to the song and the user is more enjoyable with the song with larger the number. In Table 2, the 'Song-Emotion Matrix' include each song's preference vector in different emotion. For example, song 'Opera' has the vector [0, 2, 1] in emotion type Romantic, Blue and Sadness respectively, the numeric information represents the degree of such emotion. Here the 'Song-Emotion Matrix' is the content description of the items.

$$S_{ui} = \theta_u^T x_i \tag{1}$$

To learning all the users' $\theta$ values, we need to do the following optimization:

$$min_{\theta_u} \frac{1}{2} \sum_{u=1}^{n_u} \sum_{i:r(i,u)=1} (S_{ui} - r_{ui})^2 + \frac{\lambda}{2} \sum_{u=1}^{n_u} \sum_{n=1}^{n_i} (\theta_{(n,u)})^2 \tag{2}$$

Such system has some obvious shortcomings. For the first problem, it's impossible for people to obtain all the features of items like movies, songs or hotels. Information retrieval techniques fails to extract the features of such items efficiently and accurately, and it is also time consuming and a tedious work for human beings to tag the items manually streaming of the big data nowadays. Secondly, contend based systems will only recommend items similar to what the user has purchased or rated, in other words, it is unable to recommend those items that never appeared in the user's rating history. Mutation operation was proposed to alleviate such defect [20].

### *1.2. Collaborative Filtering Recommender System*

Collaborative filtering recommenders utilize users' or products' preferences to make personal product suggestions and are prevailing in industrial recommender system such as Amazon [1], Google news [2].
Such systems perform quite satisfactory facing the challenge of big data stream in recent years. While in the same time, collaborative filtering also faces the quandary in addressing the sparse problem of e-commerce data, multiple papers proposed their solution to this problem like Associative Retrieval Techniques [3], Trust inference [17] and etc.
In Neighbourhood Collaborative Filtering Algorithm, we try to compute the similarity of users and make recommendations based on similar users. In this method, we will find the 'nearest neighbours' for each user, we assume that if two users' ratings are with high correlation, such two users must enjoy similar items and products.
In our proposed hybrid models, we use the matrix factorization based collaborative filtering algorithm which aims to find the matrix's missing items or predicting the missing entries. The details of this algorithms will be described and introduced in the next part.

### *1.3. Sequential Mining Algorithm*

We hope to introduce behavior pattern mining method to optimize the system. The sequential pattern mining algorithm can efficiently extract the frequent patterns from the database.

Traditional sequential mining algorithms such as GSP adopt a candidate sub-sequence generation-and-test approach, which will generate a huge set of candidates, and is obviously not an ideal answer for big data stream condition. Improved sequential mining algorithm such as Prefix-Span [4] adopt FP-growth strategy without candidate generation.

[5] proposed a intention prediction model to filter objectionable content for web browsers. The model use m-gram HMM to predict the user's behavior pattern and the prediction precision is comparably high. To further precisely advertise in mobile web environment, [6] proposed a method of SMAP-Tree (sequential mobile access pattern) which is similar to FP-growth strategy, such model uses sequential mining algorithm to extract behavior patterns from user's GPS logs. To build a user behavior based recommender system, [7] combined both CF and SPM (sequential pattern mining) strategies.

## Proposed system

### *1.4. Architecture of the Hybrid Model*

Fig.1 shows the system's architecture for the hybrid recommending mechanism. The workflow of the system consists of three phases: Behavior Prediction Phase, CF Phase and Recommend Phase.

We use the desensitized transaction records provided by T-mall, Alibaba which contains ten thousand users' twenty million historical purchasing data. We divide the consumer's different behaviors into 4 categories: Click, Collect, Add to Cart and Payment. The hybrid model consists of CF (collaborative filtering) and behavior prediction model. The BPM (Behavior pattern model) is in charge of payment behavior prediction work.

BPM utilize the Prefix-span algorithm to extract the most prevailing purchasing sequences from the warehouse in real time, and match the sequences with the customer's behavior pattern who is browsing or adding an item to cart.

The real time BPM will return a set of the potential purchasing behavior and the category of the purchasing item. When the recommender system's behavior monitoring part detects the users' potential purchasing tendencies, the system will fetch the user's historical behavior record from sequence database and build an item-user rating matrix (Fig. 2).

In CF phase, we use collaborative filtering method to find a set of customers whose purchased and rated items overlap the user's purchased and rated items. The algorithm generates recommendations based on a few customers who are most similar to the user. The CF method can generate the preference tendencies of the users based on their historical purchasing record. We choose top N items from the matrix as the suggested items, where N is preset parameter.

The SPM part will return the probability of the user's payment behavior's occurrence and the category of the item. If the probability returned passed the minimum preset probability, we will match the category of the item with the N items provided by the collaborative filtering model and choose those matched to recommend.

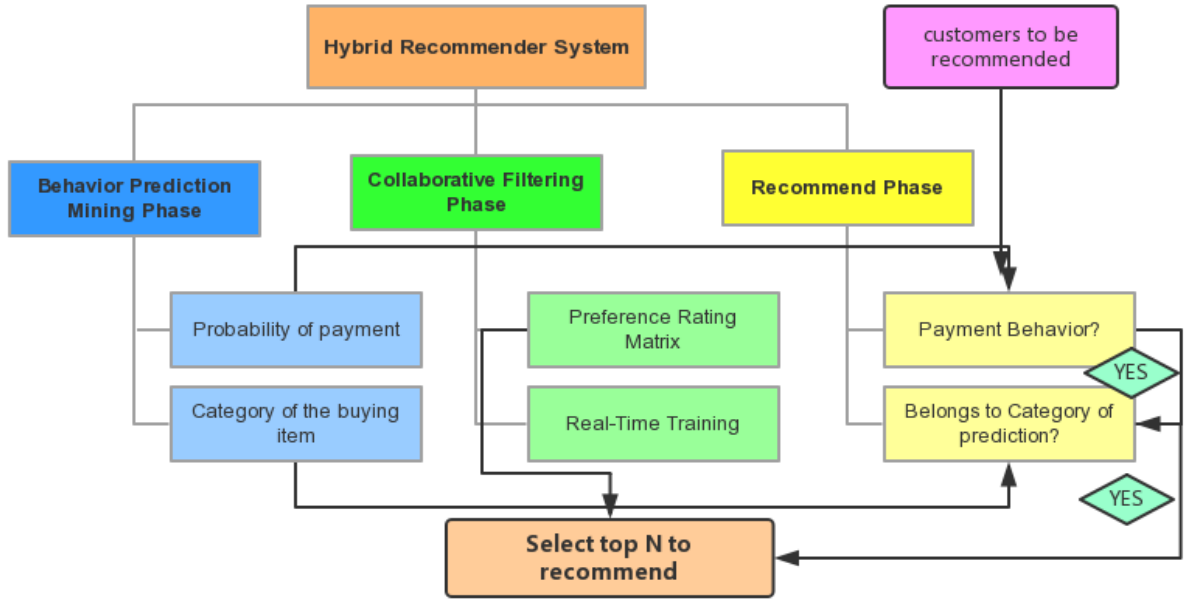

**Figure 1.** Architecture of the Hybrid Model

### 1.5. Matrix Factorization Based Collaborative Filtering

Collaborative filtering is a traditional method for products recommendation which analyzes relationships between users and inter-dependencies among products to identify new user-item associations. In order to do recommendation to users, we make two basic assumptions:
1. **User's preference to products is determined only by static features of the products.**
2. **User's preference and product features can be quantified by a linear function.**

Following the assumptions, we construct a user-item matrix, each entry contains the historical behavior of the $I^{th}$ user to $J^{th}$ product. We use number 1, 2, 3, 4 to represent click, collect, add to cart and payment respectively. We use basic matrix factorization models which map both users and items to a joint latent factor space, such that user-item interactions are modeled as inner products in that space. The next step is to factorize this matrix into two matrixes, one represent features of our products, and another represent the preferences of our users. Multiply the two matrix give back our predictions about user's preference to all products.

$$r_{ui} = q_i^T p_u \qquad (3)$$

The $r_{ui}$ represents users u's rating of item $i$, and the challenge in matrix factorization model is computing the mapping of each item and user to factor vectors $q$, $p$, $R^f$.

Since the sparseness of the user-item matrix, SVD is not an appropriate method in decompose the target matrix. We use latent factor models [8] to learn the factor vectors ($p^u$ and $q^i$), by minimizing the regularized squared error on the set of known ratings:

$$min_{p,q} \sum_{(u,i)\in \kappa} (r_{ui} - q_i^T p_u)^2 + \lambda(||q_i||^2 + ||p_u||^2) \qquad (4)$$

## 1.6. Purchasing Patterns' Extraction

Sequential pattern mining, which discovers frequent sub-sequences as patterns in a sequence database, is an important data mining problem with broad applications, including the analysis of customer purchase patterns or Web access patterns, the analysis of sequencing or time- related processes such as scientific experiments, natural disasters, and disease treatments, the analysis of DNA sequences, etc. [4].

Traditional sequential pattern mining includes GSP algorithm, Free-Span algorithm's algorithm is a priori like method which hold the principle that "all super patterns of the non-frequent pattern can not be frequent patterns, and it requires the generation of candidates thus requires large amount of storing resource. Besides that, GSP will constantly scan the database to fetch the length-l sequence which means a non-trivial cost when processing the long sequence. Apart from the deficiency of GSP algorithm, there are other crucial issues that should be taken into considerations.

1. The sequence in our SPM contains two-days' record, and the length distribution of the sequence can be shown from Fig 3. The average length is mostly distributed in length between 10 and 40 and which will require enormous computation resource.
2. The SPM should be designed to be real time functioning, which means a continuous updating to the customer's behavior sequence set once there were behavior occurrence.
3. In this hybrid system, we assume that the prevailing payment patterns are multifarious dependent variables, which means consumer's payment patterns are always changing because of discounts, advertisements, or tsunami, earthquake, and so forth. Based on factors listed above, we decide to use Prefix-Span algorithm as our pattern extraction method. Prefix-Span is a pattern growth-based approach, similar to FP-growth [4]. Prefix-Span grows longer patterns from shorter ones by dividing the search space and focusing on the subspace potentially supporting further pattern growth which requires less memory space for searching.

## 1.7. Recommending Phase

Recommend phase is the combination step of formal phases, we combine the payment behavior patterns extracted from the behavior prediction phase and the preference collected from collaborative filtering to select target items as suggestions. Fig 4. give a real example of the sequence database in which each character string represents a unique behavior and will further explained in experimental part. Let's assume character 'd' represents the payment behavior, the figure above then illustrates a payment patterns database.

In the first step, we generate the customers' real-time behavior sequences and store those sequences into database, and we call this database as candidate database. The candidate database will be scanned at a regular interval and sequence contains payment patterns will be sent to recommender system as potential purchasing sequence.

Secondly, for those potential buyers, we will generate the preference information from collaborative filtering phase which represents the preference degree toward each product. Since sequential mining phase will not only generate the payment sequence, but also the category of the target item (specific detail will be discussed in experimental part), we will choose the category matched items in preference vector to recommend.

| ITEM_ID/USERID | 100011562 | 100024529 | 100086267 | 100637858 | 100854241 |
|---|---|---|---|---|---|
| 100019569 |  | 2 | 4 |  | 1 |
| 100022999 | 1 | 1 | 2 |  |  |
| 10000003 |  | 1 |  |  |  |
| 100009489 | 3 |  | 2 | 1 | 3 |
| 100018271 |  | 1 |  |  |  |
| 100020308 |  |  | 3 |  |  |
| 100002111 | 2 | 1 |  | 1 |  |
| 100003463 |  |  | 4 |  |  |
| 10000192 | 1 | 3 |  |  |  |
| 100004450 |  |  |  |  |  |
| 100004291 |  | 4 | 1 | 1 | 4 |

USER/ITEM MATRIX

**Figure 2.** Overview of the Rating(Preference) Matrix

## Experiment data set and design

### 1.8. Data set Description

The raw data set applied in our evaluation model is an open data set provided by Session1, Alibaba's Ali Mobile Recommendation Algorithm Competition. It consists of 10,000 Alibaba customers' 12,256,906 pieces of behavior transaction data during one month on the e-commerce site. Among this data set, 480,723 kinds of products including 991 different categories are in our recommending range. For each piece of the transaction data, it reflects one customer's behavior type towards a specific product at a certain time. User's behaviors have been generalized into four types including click, collect, add-to-cart and payment behaviors, the corresponding numeric values are 1, 2, 3 and 4 respectively. Since most part of the data is closely related to user's personal information, the whole data set has been desensitized by the provider.

### 1.9. Data Pre-processing

Traditionally, rating matrix is used to perform the model training work in which ratings represent user's feedback or preference toward a specific item. For the matrix factorization based collaborative filtering, we need to transfer our raw data set into a rating matrix. Considering the values in rating matrix usually represents the preference a user possess to a specific item, so ratings in traditional models are usually considered to be explicit feedback information. In our behavior data set, we use user's different behavior types to serve as each customer's implicit ratings towards different products.

The dimension of our ratings matrix is 10,000×480,723 and is a highly sparse matrix. If we use parameter S to measure the sparseness of this matrix [19, 20],

$$S = \frac{NV}{V} \tag{5}$$

Here NV denotes the count of nonzero values and V the counts of values of this matrix, the S value of our ratings matrix is almost 1.19*e-4. Contrary to most matrix factorization methods like SVD, matrix completion via alternating least square(ALS) [21, 22, 23] is more applicable for matrix in that sparse format.

To use sequential pattern mining algorithm to extract user's purchasing patterns, we transfer the raw data set into a sequential database format. And we choose 7 days as our time span to generate behavior sequences for the following reasons.

In 1989, Prof Barbara E. Kahn [24, 25] in UCLA conducted an empirical investigation that explore the customers' shopping trip behaviors and their distributions of times between their shopping. Briefly, their analyses yielded two general conclusions about the shopping trip behaviors:
1. The distribution of times between purchases of a specific brand are significantly different. Whereas both inter-purchase times and shopping trip show evidence of 7-day (or multiples of 7 days) cycle.
2. There is evidence that these 7-day peaks in the shopping trip data are created by strong preferences for shopping on a specific day of the week rather than based on when the product is used up.

Based on such implications and conclusions, we choose one week as our behaviors' cycle. Besides that, we choose the day in which the purchasing behaviors conducted as the peak day and all behaviors before the peak day within 7 days will be concluded in such user's behavior sequence.

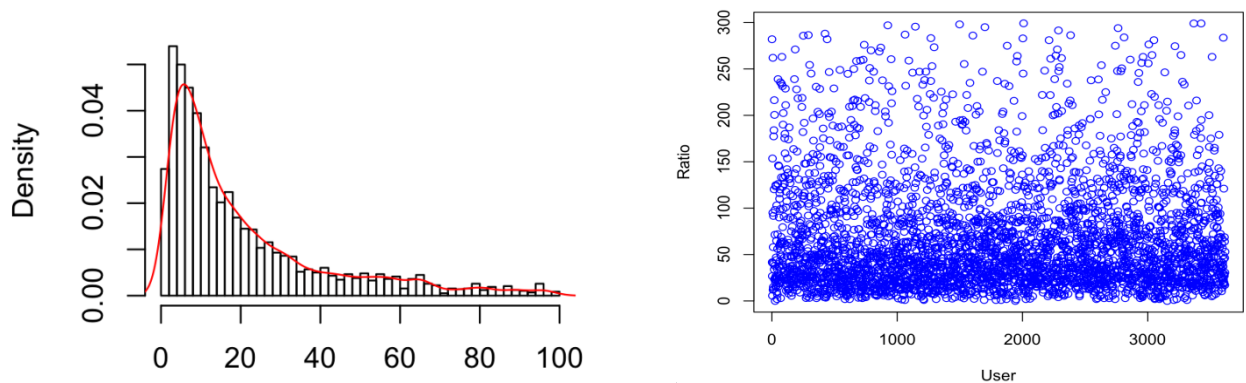

**Figure 3.** Histogram Distribution of the sequences' length

From the above picture we can see that most of the users' behavior sequences' length are within 70, which is acceptable for our model's computation.

## *1.10. Implementation of the hybrid models*

We constructed our hybrid models consists of collaborative filtering method and frequent patterns extraction algorithm and we build several groups of other traditional models to evaluate our models behaviors. We select our baseline model (BM) based of naive conditional selection because this method is usually the most common way we use in traditional recommending framework (see Table 2). We also build a nearest neighbors based collaborative filtering method (NNCF), this method investigates similar users or items, and recommendations are derived from such nearest neighbors of users/items. Matrix factorization based CF (MFCF) method tries to obtain user's and item's factor vector. For the sequential patterns mining part, algorithms GSP and SPADE were built to give a comparison to Prefix-Span algorithm. The last model is our proposed hybrid model (HM) which combined both MFCF and Prefix-Span algorithm.

**Table 1.** Models for recommendations.

| Models | Features |
|---|---|
| BM | |
| GSP +MFCF | |
| SPADE+NNCF | |
| SPADE +MFCF | |
| Prefix-Span+ MFCF(HM) | |

## *1.11. Evaluations*

The performances of this hybrid model were measured based on its ability to recommend correct items to a specific user. We adopted performance metrics consists of precision, recall, F-measurement to evaluate the results. But different to what most models, we select on single target day to deploy our prediction experiment and evaluation. Because one of the key point in our model is that we can extract correct purchasing patterns within a time span which is 7 days in our model, while actually the most prevailing purchasing patterns varied and changed quickly (see picture 2 below). For traditional collaborative models, evaluations will only consider the preferences' or rating' correctness, while for behavior tendency especially for purchasing behavior's prediction, evaluations in long time period is meaningless considering multiple behaviors could have been conducted on one single items. Besides that, items' recommendations are usually for e-commerce sites, for which precise analysis in one single day will clearly bring far more benefits and profits.

$$\text{Precision} = \frac{|\cap(predictionSetm, referenceSet)|}{|predictionSet|} \qquad (6)$$

$$\text{Recall} = \frac{|\cap(predictoinSet, referenceSet)|}{|referenceSet|} \qquad (7)$$

$$F1 = \frac{2 \times precision \times recall}{precision + recall} \qquad (8)$$

In our evaluation model, higher recall represents more fraction of items that are purchased by the user from recommendations have been successfully retrieved. In binary classification, recall is called sensitivity. So it can be looked at as the probability that a relevant item is retrieved by the recommendations. Precision value is fraction of retrieved documents that are relevant to the user. F-measurement is a measure that combines precision and recall is the harmonic mean of precision and recall .This measure is approximately the average of the two when they are close, and is more generally the square of the geometric mean divided by the arithmetic mean.

## Experimental results

### *1.12. Frequent Purchasing Behavior Patterns' Extraction*

We construct our behavior patterns' sequential database for each of the user-category pairs within 7 days. We generated 288,856 valid sequences for 10,000 users on the whole data set within 7 days. We use GSP, Spade and Prefix-span algorithms to discover the frequent patterns and retrieved 46,000, 7,000 and 10,000 patterns respectively. We also assign greater weight to patterns with higher confidence that the Top-N selection procedure will recommend those higher weight patterns firstly.

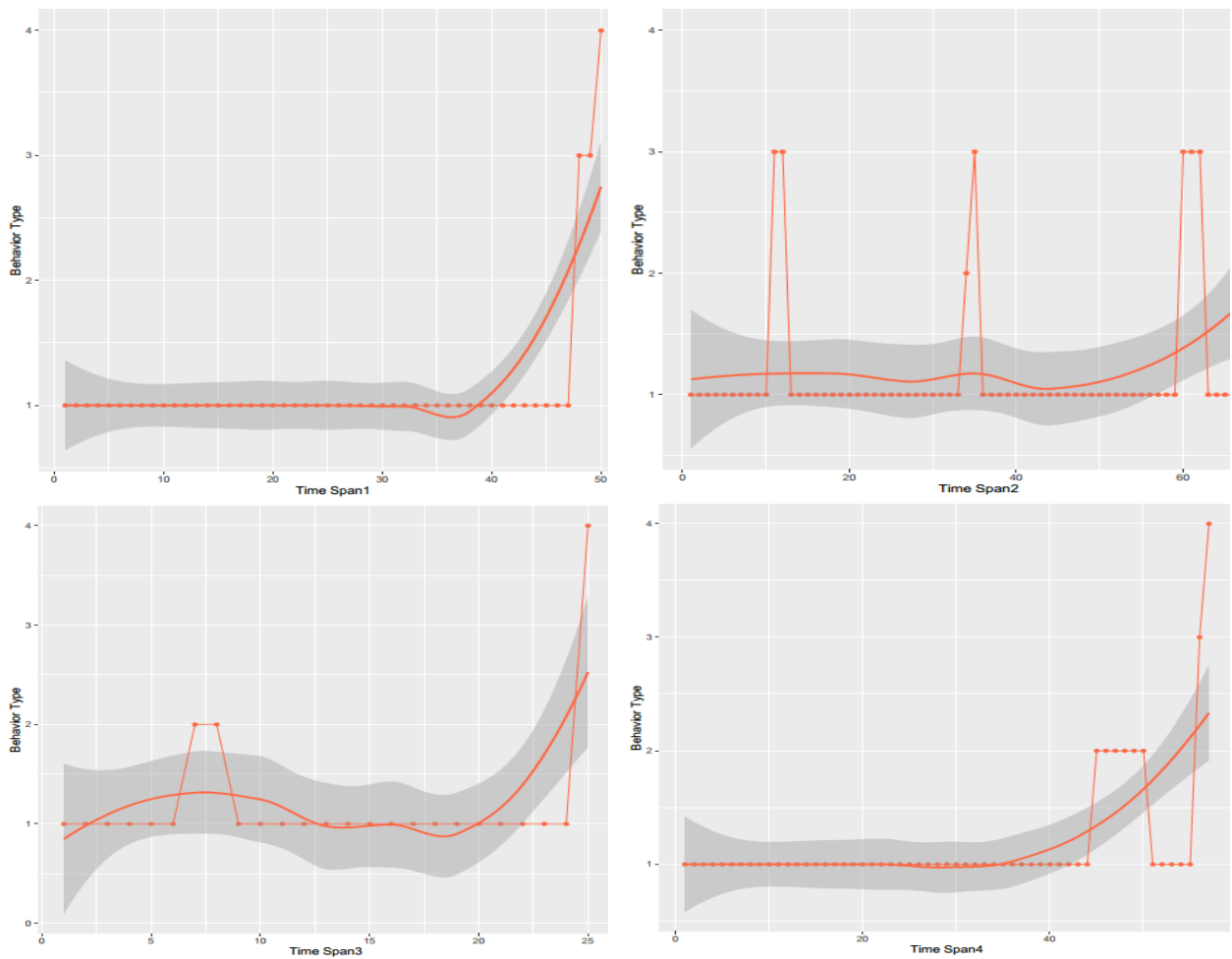

**Figure 4.** User's most prevailing purchasing patterns in different time span

The picture above is user's 4 different most prevailing purchasing patterns in on month with each time span is 7 days, and the smooth curves represent the tendency of the user's behavior. User's behaviors are described by numeric information 1, 2, 3, 4 which means click, collect, add-to-cart and payment behaviors respectively. We can deduce from that difference that customer's purchasing patterns actually varied along with the time past, but the tendencies of the same group of users who are likely to conduct purchasing behaviors remain almost the same.

## *1.13. Users' Classification Based on Statistic Feature*

Considering the variant purchasing behaviors types the different customers possess, we hold the assumption that recommendations towards different users should vary. In order to do a fine-grained prediction and recommendations, classifications on the users and items are indispensable. We firstly conduct a statistic experiment on the whole data set with the following features:

1. The user's counts of payment behaviors
2. Ratio of user's counts of browsing behaviors to counts of payment behaviors:

$$R = \frac{Exploration}{Purchase} \tag{9}$$

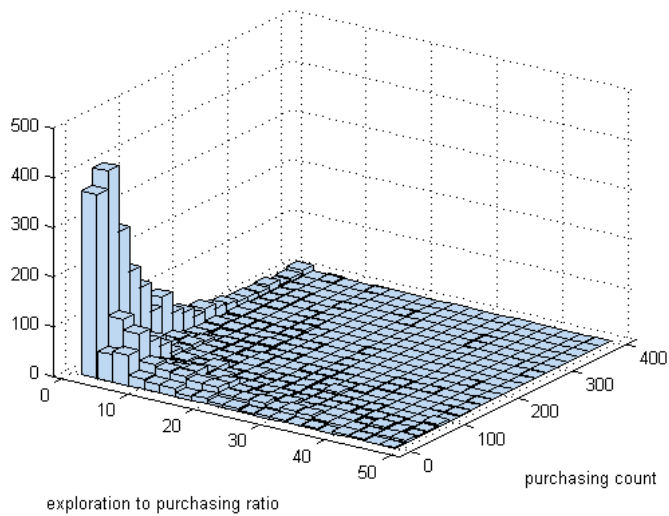

**Figure 5.** User's distribution overview on two features

Intuitively, a high exploration to purchasing ratio would indicate this user is prudent, in the sense that he would compare various different products many times before he makes a purchasing decision. For user with low exploration to purchasing ratio, he is likely to buy product without considering too much. So we expect the rule for those two kinds of users would be very different. We use the same strategy for products, namely, use the ratio of exploration count to the purchasing count as a feature to describe a specific product. Based on this feature mentioned above, we divide all users into three different groups, and we also divide all products into three different groups. So totally we have 27 groups of user – product. We expect that we would observe different patterns and prediction results for each of those groups. Secondly, we construct a nearest neighbors based collaborative filtering system and did a fine-grained prediction evaluation (See figure 4). We can clearly see from figure 4 that group 1 which stands for the users with higher exploration to purchasing ratio and purchasing counts outperforms the other groups apparently.

**Table 2.** The Threshold Of the Less Fine Item Classification

|  | Ratio of exploration to purchasing | Counts of payment behaviors |
|---|---|---|
| Item group 1 | 0-15 | 0-50 |
| Item group 2 | 15-20 | 0-50 |
| Item group 3 | 20-25 | *50-150* |
| Item group 4 | More than 25 | More than 150 |

**Table 3.** Fine-grained NNCF's Rrecommendations'performances On One Single Day

| Precision/F-measure | User group1 | | Group2 | | Goup3 | |
|---|---|---|---|---|---|---|
| Item | P | F | P | F | P | F |
| group1 | 0.258 | 0.192 | 0.120 | 0.086 | 0.085 | 0.037 |
| group2 | 0.262 | 0.195 | 0.236 | 0.225 | **0.255** | **0.167** |
| group3 | **0.359** | **0.265** | **0.279** | **0.230** | 0.199 | 0.150 |
| group4 | 0.246 | 0.255 | 0.153 | 0.152 | 0.085 | 0.072 |

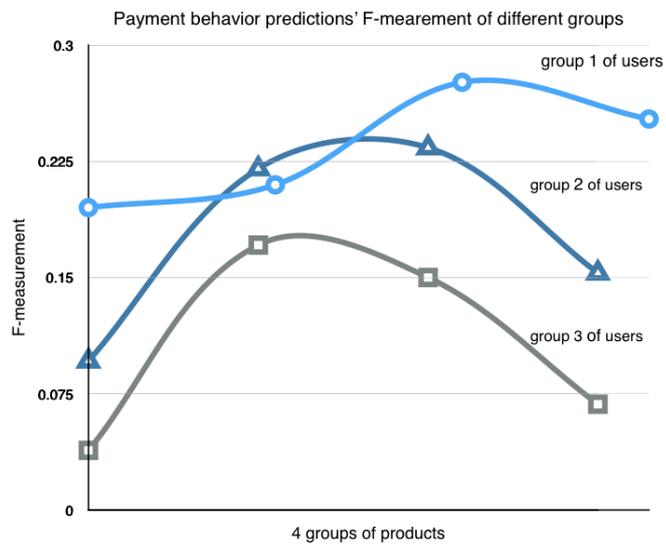

**Figure 6.** F-Measurement of NNCF in Fine-grained group

*1.14. Hybrid model of MFCF and Prefix-span's Evaluation*

Comparing the recommendation performances of different models, we conduct experiments on the selected 10,000 customers in one day. The results reported in Table 4 are based on three hybrid models and one baseline model. As shown in Table 3, the HM has higher values than the BM with respect to F-measure and precision. Therefore, behavior mining adds valuable information to the hybrid recommendation model and improve its performance.

When comparing the models' performances as respect to different groups of users, we could see clearly from Fig 6 that users with higher counts of payment behavior and lower R values outperforms than results in other groups of users.

As to the groups of items, groups 3 (which represents items been purchased comparatively frequently but with parameter R in a moderate value) possess the best performance. And we can see from such results that the classification based on behavior features differ the performances distinctively.

An intuitive explanation is that users purchase frequently and with lower R values represents a group of active people and with a more prudent process when purchasing, which can be understood as more fixed purchasing patterns during the purchasing.

It is also notable that the GSP based hybrid model's performances are not improved compared with the Spade and Prefix-Span's performances. When examining the selected patterns extracted, 46,000 valid patterns extracted by GSP in which case Spade and Prefix-Span extracted 7,000 and 10,000 respectively. The last two algorithms clearly extract the patterns more representatively. This result means that in the experimental scenario, behavior patterns are more important than preference features when placing the two types of features in a model together.

**Table 4.** Recommendation performances of different models on 3 groups of users

|        | GSP+ MCCF | | Spade+ NNCF | | Prefix-Span+ MCCF (HM) | | BM | |
|--------|-----------|-----------|---|---|---|---|---|---|
|        | Precision | F-measure | P | F | P | F | P | F |
| Group1 | 0.114 | 0.197 | 0.198 | *0.253* | **0.286** | **0.276** | 0.121 | 0.099 |
| Group2 | 0.062 | 0.115 | 0.179 | **0.234** | **0.179** | **0.234** | 0.081 | 0.089 |
| Group3 | 0.026 | 0.050 | 0.090 | 0.150 | **0.090** | **0.151** | 0.067 | 0.094 |
| Overall | *0.049* | *0.093* | 0.128 | 0.178 | **0.153** | **0.196** | *0.073* | *0.0812* |

## 1.15. Precision and F-measure curve

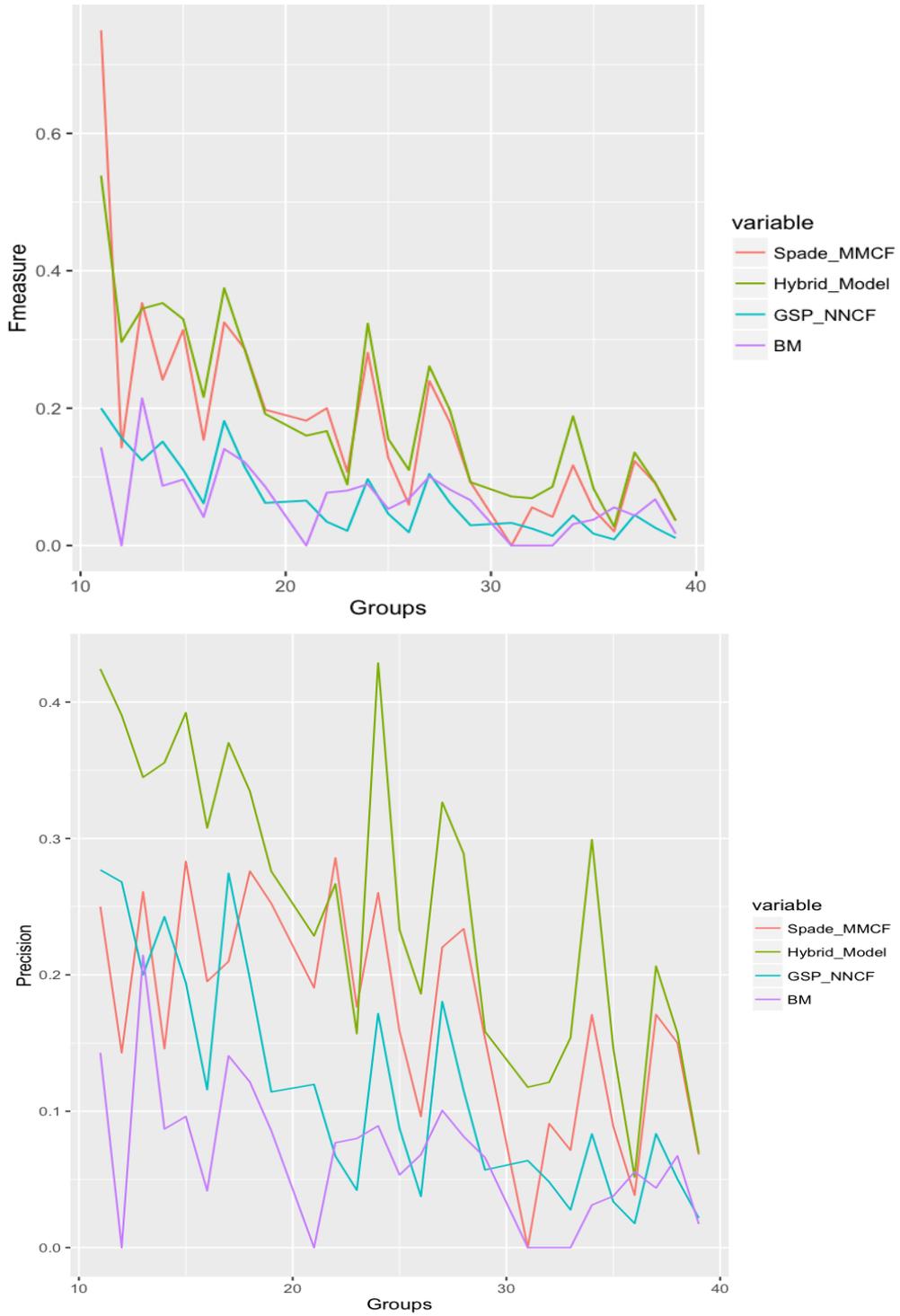

**Figure 7.** Fine-grained Experiment Precision and F-Measure with Different Groups

# Discussion and conclusion

In this paper, we propose a new hybrid recommender system based on behavior mining and it has been proved to performs better than the now existing traditional models. We carried out our evaluation part on a precise prediction task in one day which indicates the specific user and products should be predicted from their historical transaction record. Intuitively, traditional recommender systems try to extract user' potential preference by search similar users, products and are quite applicable in this kind of prediction job. Intuitively, we need to design a hybrid system to be more personalized that will consider each customer's behavior tendency. In our experiment, we construct users' behavior sequences and extract those frequent purchasing sequences to further predict users' probability to conduct purchasing behaviors. Moreover, we also take the advantage of the matrix factorization based CF algorithm which has been verified to posses the best performances in the analysis of users' purchasing preference. By combining such two methods, this hybrid model can predict far better than each of the single model.

Despite its ability in precise prediction work, some possible extension could be as follows. Firstly, limited by the raw dataset provided by Alibaba, we could not obtain more personalized information like location, context information of products and so forth. Thus the purchasing patterns are not the only aspect of information we are able to utilize theoretically. More behavior information should be considered and proposed in the future's research and work. Secondly, recommending models for different groups of users or items should vary. In this experimental scenario, we verify the users with more fixed purchasing patterns are far more suitable for the proposed hybrid models. As for the users that are less active and with transitory purchasing patterns, we should propose proper improved methods.